# Design Guidelines for Domain Specific Languages

Gabor Karsai
Institute for Software
Integrated Systems
Vanderbilt University
Nashville, USA

Holger Krahn
Software Engineering Group
Department of Computer Science
RWTH Aachen, Germany

Claas Pinkernell
Software Engineering Group
Department of Computer Science
RWTH Aachen, Germany

Bernhard Rumpe
Software Engineering Group
Department of Computer Science
RWTH Aachen, Germany

Martin Schindler
Software Engineering Group
Department of Computer Science
RWTH Aachen, Germany

Steven Völkel
Software Engineering Group
Department of Computer Science
RWTH Aachen, Germany

## ABSTRACT

Designing a new domain specific language is as any other complex task sometimes error-prone and usually time consuming, especially if the language shall be of high-quality and comfortably usable. Existing tool support focuses on the simplification of technical aspects but lacks support for an enforcement of principles for a good language design. In this paper we investigate guidelines that are useful for designing domain specific languages, largely based on our experience in developing languages as well as relying on existing guidelines on general purpose (GPLs) and modeling languages. We defined guidelines to support a DSL developer to achieve better quality of the language design and a better acceptance among its users.

## 1. INTRODUCTION

Designing a new language that allows us to model new technical properties in a simpler and easier way, describe or implement solutions, or to describe the problem resp. requirements in a more concise way is one of the core challenges of computer science. The creation of a new language is a time consuming task, needs experience and is thus usually carried out by specialized language engineers. Nowadays, the need for new languages for various growing domains is strongly increasing. Fortunately, also more sophisticated tools exist that allow software engineers to define a new language with a reasonable effort. As a result, an increasing number of DSLs (Domain Specific Languages) are designed to enhance the productivity of developers within specific domains. However, these languages often fit only to a rather specific domain problem and are neither of the quality that they can be used by many people nor flexible enough to be easily adapted for related domains.

During the last years, we developed the frameworks MontiCore [13] and GME [2] which support the definition of domain specific languages. Using these frameworks we designed several DSLs for a variety of domains, e.g., a textual version of UML/P notations [17] and a language based on function nets in the automotive domain [5]. We experienced that the design of a new DSL is a difficult task because different people have a varying perception of what a "good" language actually is.

This of course also depends on the taste of the developer respectively the users, but there are a number of generally acceptable guidelines that assist in language development, making it more a systematic, methodological task and less an intellectual ad-hoc challenge. In this paper we summarize, categorize, and amend existing guidelines as well as add our new ones assuming that they improve design and usability of future DSLs.

In the following we present general guidelines to be considered for both textual and graphical DSLs with main focus is on the former. The guidelines are discussed sometimes using examples from well-known programming languages or mathematics, because these languages are known best. Depending on the concrete language and the domain these guidelines have to be weighted differently as there might be different purposes, complexity, and number of users of the resulting language. For example, for a rather simple configuration language used in only one project a timely realization is usually more important than the optimization of its usability. Therefore, guidelines must be sometimes ignored, altered, or enforced. Especially quality-assurance guidelines can result in an increased amount of work.

While we generally focus in our work on DSLs that are specifically dedicated to modeling aspects of (software) systems, we believe that these guidelines generally hold for any DSL that embeds a certain degree of complexity.

### 1.1 Literature on Language Design

For programming languages, design guidelines have been intensively discussed since the early 70s. Hoare [8] introduced simplicity, security, fast translation, efficient object code, and readability as general criteria for the design of good languages. Furthermore, Wirth [22] discussed several guidelines for the design of languages and corresponding compilers. The rationale behind most of the guidelines and hints of both articles can be accepted as still valid today, but the technical constraints have changed dramatically since the 70s. First of all, computer power has increased significantly. Therefore, speed and space problems have become less important. Furthermore, due to sophisticated tools (e.g., parser generators) the implementation of accompanying tools is often not a necessary part of the language development any more. Of course, both articles concentrate on programming languages and do not consider the greater variety of domain specific languages.

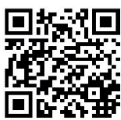

More recently, authors have also discussed the design of domain specific modeling languages. General principles for modeling language design were introduced in [14]. These include simplicity, uniqueness, consistency, and scalability, on which we will rely later. However, the authors did not discuss how these higher level principles can be achieved. In [12] certain aspects of the DSL development are explained and some guidelines are introduced. More practical guidelines for implementing DSLs are given in [10]. These focus on how to identify the necessary language constructs to generate full code from models. The authors explain how to provide tool support with the MetaEdit+ environment. [20] explains 12 lessons learned from DSL experiments that can help to improve a DSL. Although more detailed discussions on explicit guidelines are missing, these lessons embed documented empirical evidence – a documentation that many other discussions, including ours do not have. In [16] the authors introduce a toolset which supports the definition of DSLs by checking their consistency with respect to several objectives. Language designers can select properties of their DSL to be developed and the system automatically derives other design decisions in order to gain a consistent language definition. However, the introduced criteria cover only a subset of the decisions to be made and hence, cannot serve as the only criteria for good language design. Quite the contrary, to our experience many design guidelines cannot be translated in automatic measures and thus cannot be checked by a tool.

## 1.2 Categories of DSL Design Guidelines

The various design guidelines we will discuss below, can be organized into several categories. Essentially, these guidelines describe techniques that are useful at different activities of the language development process, which range from the domain analysis to questions of how to realize the DSL to the development of an abstract and a concrete syntax including the definition of context conditions. An alignment of guidelines with the language development activities and the developed artifacts has the advantage that a language designer can concentrate on the respective subset of the guidelines at each activity. This should help identifying and realizing the desired guidelines. Therefore, we decided for a development phase oriented classification and identified the following categories:

**Language Purpose** discusses design guidelines for the early activities of the language development process.

**Language Realization** introduces guidelines which discuss how to implement the language.

**Language Content** contains guidelines which focus on the elements of a language.

**Concrete Syntax** concentrates on design guidelines for the readable (external) representation of a language.

**Abstract Syntax** concentrates on design guidelines for the internal representation of a language.

For each of these categories we will discuss the design guidelines we found useful. Please be aware that the subsequently discussed guidelines sometimes are in conflict with each other and the language developer sometimes has to balance them accordingly. Additionally, semantics is explicitly not listed as a separate step as it should be part of the entire development process and therefore has an influence on all of the categories above.

## 2. DSL DESIGN GUIDELINES

### 2.1 Language Purpose

Language design is not only influenced by the question of what it needs to describe, but equally important what to do with the language. Therefore, one of the first activities in language design is to analyze the aim of the language.

**Guideline 1:** *"Identify language uses early."* The language defined will be used for at least one task. Most common uses are: documentation of knowledge (only) and code generation. However, there are a lot more forms of usage: definition or generation of tests, formal verification, automatic analysis of various kinds, configuration of the system at deployment- or run-time, and last but increasingly important, simulation.

An early identification of the language uses have strong influence on the concepts the language will allow to offer. Code generation for example is not generally feasible when the language embeds concepts of underspecification (e.g., non-deterministic Statecharts). Even if everything is designed to be executable, there are big differences regarding the overhead necessary to run certain kinds of models. If efficient execution on a small target machine is necessary (e.g., mobile or car control device) then high-level concepts must be designed for optimized code generations. For simulation and validation of requirements however, efficiency plays a minor role.

**Guideline 2:** *"Ask questions."* Once the uses of a language have been identified it is helpful to embed these forms of language uses into the overall software development process. People/roles have to be identified that develop, review, and deploy the involved programs and models. The following questions are helpful for determining the necessary decisions: Who is going to model in the DSL? Who is going to review the models? When? Who is using the models for which purpose?

Based thereon, the question after whether the language is too complex or captures all the necessary domain elements can be revisited. In particular, appropriate tutorials for the DSL users in their respective development process should now be prepared.

**Guideline 3:** *"Make your language consistent."* DSLs are typically designed for a specific purpose. Therefore, each feature of a language should contribute to this purpose, otherwise it should be omitted. As an illustrative example we consider a platform independent modeling language. In this language, all features should be platform independent as well. This design principle was already discussed in [14].

### 2.2 Language Realization

When starting to define a new language, there are several options on how to realize it. One can implement the DSL from scratch or reuse and extend or reduce an existing language, one can use a graphical or a textual representation, and so on. We have identified general hints which have to be taken into account for these decisions.

**Guideline 4:** *"Decide carefully whether to use graphical or textual realization."* Nowadays, it is common to use tools supporting the design of graphical DSLs such as the Eclipse Modeling Framework (EMF) or MetaEdit+. On the other hand, there exist sophisticated tools and frameworks like MontiCore or xText for text-based modeling. As described in [6], there are a number of advantages and disadvantages for both approaches. Textual representations for example usually have the advantage of faster development and are platform and tool independent whereas graphical models provide a better overview and ease the understanding of models. Therefore, advantages and disadvantages have to be weighted and matched against end users' preferences in order to make a substantiated decision for one of the realizations. From this point on, a more informed decision can be made for a concrete tool to realize the language based on their particular features and the intended use of the language. Comparisons can be found in [21] or [3].

**Guideline 5:** *"Compose existing languages where possible."* The development of a new language and an accompanying toolset is a labor-intensive task. However, it is often the case that existing languages can be reused, sometimes even without adaptation. A good example for language reuse is OCL: it can be embedded in other languages in order to define constraints on elements expressed in the hosting language.

The most general and useful form of language reuse is thus the unchanged embedding of an existing language into another language. A more sophisticated approach is to have predefined holes in a host language, such that the definition of a new language basically consists of a composition of different languages. For textual languages this compositional style of language definitions is well understood and supported by sophisticated tools such as [11] which also assists the composition of appropriate tools.

However, according to the seamlessness principle [14], the concepts of the languages to be composed need to fit together. In the UML, the object oriented paradigm underlies both class diagrams and Statecharts which therefore fit well together. Additionally, when composing languages care must be exercised to avoid confusion: similar constructs with different semantics should be avoided.

**Guideline 6:** *"Reuse existing language definitions."* If the language cannot be simply composed from some given language parts, e.g., by language embedding as proposed in guideline 5, it is still a good idea to reuse existing language definitions as much as possible. In [18] more possible realization strategies, such as language extension or language specialization are analyzed. This means, taking the definition of a language as a starter to develop a new one is better than creating a language from scratch. Both the concrete and the abstract syntax will benefit from this form of reuse. The new language then might retain a look-and-feel of the original, thus allowing the user to easily identify familiar notations. Looking at the abstract syntax of existing languages, one can identify "language pattern" (quite similar to design pattern), which are good guidelines for language design. For example, expressions, primary expressions, or statements have quite a common pattern in all languages.

Only if there is no existing language/notation or the disadvantages do not allow using the strategies mentioned above, a standalone realization should be considered. The websites of parser generators like Antlr [1] or Atlantic Zoo [19] are a good starting point for reusing language definitions.

**Guideline 7:** *"Reuse existing type systems."* A DSL used for software development often comprises and even extends either a property language such as OCL or an implementation language such as Java. As described in [8], the design of a type system for such a language is one of the hardest tasks because of the complex correlations of name spaces, generic types, type conversions, and polymorphism.

Furthermore, an unconventional type system would be hard for users to adopt as well. Therefore, a language designer should reuse existing type systems to improve comprehensibility and to avoid errors that are caused by misinterpretations in an implementation. Furthermore, it is far more economical to use an existing type system, than developing a new one as this is a labor intensive and error-prone task. A well-documented object-oriented type system can be tailored to the needs of the DSL or even an implemented reusable type system can be used (e.g. [4]).

## 2.3 Language Content

One main activity in language development is the task of defining the different elements of the language. Obviously, we cannot define in general which elements should be part of a language as this typically depends on the intended use. However, the decisions can be guided by some basic hints we propose in this Section.

**Guideline 8:** *"Reflect only the necessary domain concepts."* Any language shall capture a certain set of domain artifacts. These domain artifacts and their essential properties need to be reflected appropriately in the language in a way that the language user is able to express all necessary domain concepts. To ensure this, it is helpful to define a few models early to show how such a reflection would look like. These models are a good basis for feedback from domain experts which helps the developer to validate the language definition against the domain. However, when designing a language not all domain concepts need to be reflected, but only those that contribute to the tasks the language shall be used for.

**Guideline 9:** *"Keep it simple."* Simplicity is a well known criterion which enhances the understandability of a language [8, 14, 22]. The demand for simplicity has several reasons. First, introducing a new language in a domain produces work in developing new tools and adapting existing processes. If the language itself is complex, it is usually harder to understand and thus raises the barrier of introducing the language. Second, even when such a language is successfully introduced in a domain, unnecessary complexity still minimizes the benefit the language should have yielded. Therefore, simplicity is one of the main targets in designing languages. The following more detailed Guidelines 10, 11, and 12 will show how to achieve simplicity.

**Guideline 10:** *"Avoid unnecessary generality."* Usually, a domain has a finite collection of concepts that should be reflected in the language design. Statements like "maybe we can generalize or parameterize this concept for future changes in the domain" should be avoided as they unneces-sarily complicate the language and hinder a quick and successful introduction of the DSL in the domain. Therefore, this guideline can also be defined as "design only what is necessary".

**Guideline 11:** *"Limit the number of language elements."* A language which has several hundreds of elements is obviously hard to understand. One approach to limit the number of elements in a language for complex domains is to design sublanguages which cover different aspects of the systems. This concept is, e.g., employed by the UML: different kinds of diagrams are used for special purposes such as structure, behavior, or deployment. Each of them has its own notation with a limited number of concepts.

A further possibility to limit the number of elements of a language is to use libraries that contain more elaborated concepts based on the concepts of the basic language and that can be reused in other models. Elements which were previously defined as part of the language itself can then be moved to a model in the library (compare, e.g., I/O in Pascal vs. C++). Furthermore, users can extend a library by their own definitions and thus, can add more and more functionality without changing the language structure itself. Therefore, introducing a library leads to a flexible, extensible, and extensive language that nevertheless is kept simple. On the other hand, a language capable of library import and definition of those elements must have a number of appropriate concepts embedded to enable this (e.g., method and class definitions, modularity, interfaces - whatever this means in the DSL under construction). This principle has successfully been applied in GPL design where the languages are usually small compared to their huge standard libraries.

**Guideline 12:** *"Avoid conceptual redundancy."* Redundancy is a constant source of problems. Having several concepts at hand to describe the same fact allows users to model it differently. The case of conceptual richness in C++ shows that coding guidelines then usually forbid a number of concepts. E.g., the concept of classes and structs is nearly identical, the main difference is the default access of members which is `public` for structs and `private` for classes. Therefore, classes and structs can be used interchangeably within C++ whereas the slight difference might be easily forgotten. So, it should be generally avoided to add redundant concepts to a language.

**Guideline 13:** *"Avoid inefficient language elements."* One main target of domain specific modeling is to raise the level of abstraction. Therefore, the main artifacts users deal with are the input models and not the generated code. On the other hand, the generated code is necessary to run the final system and more important, the generated code determines significant properties of the system such as efficiency. Hence, the language developer should try to generate efficient code.

Furthermore, efficiency of a model should be transparent to the language user and therefore should only depend on the model itself and not on specific elements used within the model. Elements which would lead to inefficient code should be avoided already during language design so that only the language user is able to introduce inefficiency [8]. For example, in Java there is no operator to get all instances of one class as this would increase memory usage and operating time significantly. However, this functionality can be implemented by a Java user if needed.

## 2.4 Concrete Syntax

Concrete syntax has to be chosen well in order to have an understandable, well structured language. Thus, we concentrate on the concrete syntax first and will deal with the abstract syntax later.

**Guideline 14:** *"Adopt existing notations domain experts use."* As [20] says, it is generally useful to adopt whatever formal notation the domain experts already have, rather than inventing a new one.

Computer experts and especially language designers are usually very practiced in learning new languages. On the contrary, domain experts often use a language for a longer time and do not want to learn a new concrete syntax especially when they already have a notation for a certain problem. As already mentioned, it is often the case that the introduction of a DSL makes new tools and modified processes necessary. Inventing a new concrete syntax for given concepts would raise the barrier for domain experts. Thus, existing notations should be adopted as much as possible. E.g., queries within the database domain should be defined with SQL instead of inventing a new query language. Even if queries are only part of a new language to be defined SQL could be embedded within the new language.

In case a suitable notation does not already exist, the new language should be adopted as close as possible to other existing notations within the domain or to other common used languages. A good example for commonly accepted languages are mathematical notations like arithmetical expressions [8].

**Guideline 15:** *"Use descriptive notations."* A descriptive notation supports both learnability and comprehensibility of a language especially when reusing frequently-used terms and symbols of domain or general knowledge. To avoid misinterpretation it is highly important to maintain the semantics of these reused elements. For instance, the sign "+" usually stands for addition or at least something semantically similar to that whereas commas or semicolons are interpreted as separators. This applies to keywords with a widely-accepted meaning as well. Furthermore, keywords should be easily identifiable. It is helpful to restrict the number of keywords to a few memorizable ones and of course, to have a keyword-sensitive editor.

A good example for a descriptive notation is the way how special character like Greek letters are expressed in Latex. Instead of using a Unicode-notation each letter can be expressed by its name (`\alpha` for $\alpha$, `\beta` for $\beta$, and so on).

**Guideline 16:** *"Make elements distinguishable."* Easily distinguishable representations of language elements are a basic requirement to support understandability. In graphical DSLs, different model elements should have representations that exhibit enough syntactic differences to be easily distinguishable. Different colors as the only criteria may be counterproductive, e.g., when printed in black and white. In textual languages usually keywords are used in order to separate kinds of elements. These keywords have to be placed in appropriate positions of the concrete syntax, as otherwise readers need to start backtracking when "parsing" the text [8, 22]. The absence of keywords is often based on efficiency for the writer. But this is a very weak reason because models are much more often read than written and therefore to be designed from a readers point of view.

**Guideline 17:** *"Use syntactic sugar appropriately."* Languages typically offer syntactic sugar, i.e., elements which do not contribute to the expressiveness of the language. Syntactic sugar mainly serves to improve readability, but to some extent also helps the parser to parse effectively. Keywords chosen wisely help to make text readable. Generally, if an efficient parser cannot be implemented, the language is probably also hard to understand for human readers.

However, an overuse of the addition of syntactic sugar distracts, because verbosity hinders to see the important content directly. Furthermore, it should be kept in mind that several forms of syntactic sugar for one concept may hinder communication as different persons might prefer different elements for expressing the same idea.

Nevertheless the introduction of syntactic sugar can also improve a language, e.g., the enhanced for-statement in Java 5 is widely accepted although it is conceptually redundant to a common for-statement. This is a conflict to guideline 12, but the frequency of occurrence of common for-statements in Java legitimates a more effective alternative of this notation.

**Guideline 18:** *"Permit comments."* Comments on model elements are essential for explaining design decisions made for other developers. This makes models more understandable and simplifies or even enables collaborative work. So a widely accepted standard form of grouped comments, like `/* ... */`, and line comments, like `// ...` for textual languages or text boxes and tooltips for graphical languages should be embedded.

Furthermore, specially structured comments can be used for further documentation purposes as generating HTML-pages like Javadoc. In [8] it is mentioned that the "purpose of a programming language is to assist in the documentation of programs". Therefore we recommend that every DSL should allow a user to generally comment at various parts of the model. If desired, the language may even contain the definition of a comment structure directly, thus enforcing a certain style of documentation.

**Guideline 19:** *"Provide organizational structures for models."* Especially for complex systems the separation of models in separate artifacts (files) is inevitable but often not enough as the number of files would lead to an overflowed model directory. Therefore, it is desirable to allow users to arrange their models in hierarchies, e.g., using a package mechanism similar to Java and store them in various directories.

As a consequence, the language should provide concepts to define references between different files. Most commonly "`import`" is used to refer to another name space. Imports make elements defined in other DSL artifacts visible, while direct references to elements in other files usually are expressed by qualified names like "`package.File.name`". Sometimes one form of import isn't enough and various relations apply which have to be reflected in the concrete syntax of the language.

**Guideline 20:** *"Balance compactness and comprehensibility."* As stated above, usually a document is written only once but read many times. Therefore, the comprehensibility of a notation is very important, without too much verbosity. On the other hand, the compactness of a language is still a worthwhile and important target in order to achieve effectiveness and productivity while writing in the language. Hence a short notation is more preferable for frequently used elements rather than for rarely used elements.

**Guideline 21:** *"Use the same style everywhere."* DSLs are typically developed for a clearly defined task or viewpoint. Therefore, it is often necessary to use several languages to specify all aspects of a system. In order to increase understandability the same look-and-feel should be used for all sublanguages and especially for the elements within a language. In this way the user can obtain some kind of intuition for a new language due to his knowledge of other ones. For instance, it is hardly intuitive if curly braces are used for combining elements in one language and parentheses in another. Additionally, a general style can also assist the user in identifying language elements, e.g., if every keyword consists of one word and is written in lower case letters.

A conflicting example is the embedment of OCL. One the one hand it is possible to adapt the OCL syntax to the enclosing language to provide the same syntactic style in both languages. On the other hand different OCL styles impede the comprehensibility of OCL, what endorses the use of a standard OCL syntax.

**Guideline 22:** *"Identify usage conventions."* Preferably not every single aspect should be defined within the language definition itself to keep it simple and comprehensible (see guideline 11). Furthermore, besides syntactic correctness it is too rigid to enforce a certain layout directly by the tools. Instead, usage conventions can be used which describe more detailed regulations that can, but need not be enforced.

In general, usage conventions can be used to raise the level of comprehensibility and maintainability of a language. The decision, whether something goes as a usage convention or within a language definition is not always clear. So, usage conventions must be defined in parallel to the concrete syntax of the language itself. Typical usage conventions include notation of identifiers (uppercase/lowercase), order of elements (e.g. attributes before methods), or extent and form of comments. A good example for code conventions for a programming language can be found in [9].

## 2.5 Abstract Syntax

**Guideline 23:** *"Align abstract and concrete syntax."* Given the concrete syntax, the abstract syntax and especially its structure should follow closely to the concrete syntax to ease automated processing, internal transformations and also presentation (pretty printing) of the model.

In order to align abstract and concrete syntax three main principles apply: First, elements that differ in the concrete syntax need to have different abstract notations. Second, elements that have a similar meaning can be internally represented by reusing concepts of the abstract syntax (usually through subclassing). This is more a semantics-based decision than a structurally based decision. Third, the abstract notation should not depend on the context an element is used in but only on the element itself. A pretty bad example for context-dependent notations is the use of "=" as assignment in OCL-statements (let-construct) and as equality in OCL-expressions. Here, the semantics obviously differs whilst the syntax is equal.

Furthermore, the use of a transformation engine usually also requires an understanding of the internal structure of a language, which is related to the abstract syntax. Therefore, the user to some extent is exposed to the internal structure of the language and hence needs an alignment between his concrete representations and the abstract syntax, where the transformations operate on.

Alignment of both versions of syntax and the seamlessness principle discussed in [14] assures that it is possible to map abstractions from a problem space to concrete realizations in the solution space. For a domain specific language the domain is then reflected as directly as possible without much bias, e.g., of implementation or executability considerations.

**Guideline 24:** *"Prefer layout which does not affect translation from concrete to abstract syntax."* A good layout of a model can be used to simplify the understanding for a human reader and is often used to structure the model. Nevertheless, a layout should be preferred which does not have any impact on the meaning of the model, and thus, does not affect the translation of the concrete to the abstract syntax and the semantics. As an example, this is the case for computer languages where the program structure is achieved by indentation. From a practical point of view, line separators, tabs, and spaces are often treated differently depending on editors and platforms and are usually difficult to distinguish by a human reader. If these elements gain a meaning, developers have to be much more cautious and a collaborative development requires more effort. For graphical languages a well-known bad example is the twelve o'clock semantics in Stateflow [7] where the order of the placement of transitions can change the behavior of the Statechart. To simplify the usage of DSLs, we recommend that the layout of programs doesn't affect their semantics.

**Guideline 25:** *"Enable modularity."* Nowadays, systems are very complex and thus, hard to understand in their entirety. One main technique to tackle complexity is modularization [15] which leads to a managerial, flexible, comprehensible, and understandable infrastructure. Furthermore, modularization is a prerequisite for incremental code generation which in turn can lead to a significant improvement of productivity. Therefore, the language should provide a means to decompose systems into small pieces that can be separately defined by the language users, e.g., by providing language elements which can be used in order to reference artifacts in other files.

**Guideline 26:** *"Introduce interfaces."* Interfaces in programming languages provide means for a modular development of parts of the system. This is especially important for complex systems as developers may define interfaces between their parts to be able to exchange one implementation of an interface with another which significantly increases flexibility. Furthermore, the introduction of interfaces is a common technique for information hiding: developers are able to change parts of their models and can be sure that these changes do not affect other parts of the system when the interface does not change. Therefore, we recommend that a DSL should provide an interface concept similar to the interfaces of known programming languages.

One example of interfaces are visibility modifiers in Java. They provide a means to restrict the access to members in a simple way. Another common example are ports, e.g., in composite structure diagrams, which explicitly define interaction points and specify services they provide or need, thus declaring a more detailed interface of a part of a system.

## 3. DISCUSSION

In the previous sections we introduced and categorized a bundle of guidelines dedicated to different language artifacts and development phases. Some of them already contained notes on relationships with other guidelines and trade-offs between them, and some of them briefly discussed their importance in different project settings. However, the following more detailed discussion shall help to identify possible conflicting guidelines and their reasons and gives hints on decision criteria.

The most contradicting point is reuse of existing artifacts versus the implementation of a language from scratch (cf. No. 5, 6, and 7). The main reason for the reuse of a language or a type system is that it can significantly decrease development time. Furthermore, existing languages often provide at least an initial level of quality. Thus, some of the guidelines, e.g., guidelines which target at consistency (e.g., No. 21) or claim modularity (e.g., No. 25), are met automatically. However, reusing existing languages can hinder flexibility and agility as an adaption may be hard to realize if not impossible. The same ideas apply to an improvement of the reused language itself (e.g., to meet guidelines which were not respected by the original language): the implementation of a single guideline may require a significant change of the language. Another important point is that this approach may influence the satisfiability of other guidelines. One example is No. 14 which suggests the reuse of existing notations of the domain. In case there are no languages which are similar to these notations, this guideline and language reuse are obviously contradicting. Furthermore, combining several existing languages may introduce conceptual inconsistencies, such as different styles or different underlying type systems which have to be translated into each other (cf., No. 5).

Implementing a new language from scratch in turn permits a high degree of freedom, agility, and flexibility. In this case, some guidelines can be realized more easily than in the case of reuse. However, these advantages are not for free: designing concrete and abstract syntax, context conditions, and a type system are time- and cost-intensive task. To summarize, a decision whether to reuse existing languages or to implement a new one is one of the most important and critical decisions to be made.

Another important point which was already mentioned in the introduction is that some of the presented guidelines have to be weighted according to the project settings, to the form of use, etc. One example is the expected size of the languages instances. Some DSLs serve as configuration languages and thus, typical instances consist of a small amount of lines only. Other DSLs are used to describe complex systems leading to huge instances. In the former case guidelines which target at compositionality or claim references between files (e.g., No. 19 and 25) have nearly no validity whereas in the latter example these guidelines are of high importance. However, not only the expected size of the instances can influence the weight of guidelines. Another important aspect is the intended usage of the language. Sometimes DSLs are not executable; they are designed for documentation only. In these cases, the guideline which demands to avoid inefficient elements in the language (No. 13) is of course not meaningful. However, for languages which are translated into running code, this is of high importance.

A last point we want to discuss here are the costs induced by applying the guidelines. Some of them can be implemented easily and straightforward (e.g., distinguishability of elements or permitting comments, No. 16 and 18) whilst others require a significant amount of work (e.g., introduction of references between files including appropriate resolution mechanisms and symbol tables, No. 19). Of course, especially guidelines whose implementation is cost intensive have to be matched against project settings as described above. For small DSLs such guidelines should be ignored instead as the cost will often not amortize the improvements. However, from our experiences DSLs are often subject to changes. While growing these guidelines become more and more important. The main problem which emerges in these cases is that adding new things to a grown language (e.g., modularity) is typically more difficult and time-consuming than it would have been at the beginning. Therefore, analyzing the domain and usage scenarios as described in Guidelines 1 and 2 can prevent those unnecessary costs.

## 4. CONCLUSION

In this paper 26 guidelines have been discussed that should be considered while developing domain specific languages. To our experience this set of guidelines is a good basis for developing a language. For space reasons, we restricted ourselves to guidelines for designing the language itself. Other guidelines are needed for successfully integrating DSLs in a software development process, deploying it to new users, and evolving the syntax and existing models in a coherent way.

In general, a guideline should not be followed closely, but many of them are proposals as to what a language designer should consider during development. Some of the guidelines have to be discussed in certain domains, because they might not have the same relevance and as discussed many guidelines contradict each other and the language developer has to balance them appropriately.

But generally, the consideration of explicitly formulated guidelines is improving language design. We also think that it is worthwhile to develop much more detailed sets of concrete instructions for particular DSLs. We currently focus on textual languages in the spirit of Java.

Although we have compiled this list from literature and our own experience, we are sure that this list is not complete and has to be extended constantly. In addition, guidelines might change during time as developers gather more experience, tools become more elaborate, and taste changes. Maybe some guidelines are not relevant anymore in a few years, as some guidelines from the 1970's are less important today.

*Acknowledgment:* The work presented in this paper is partly undertaken in the MODELPLEX project. MODELPLEX is a project co-funded by the European Commission under the "Information Society Technologies" Sixth Framework Programme (2002-2006). Information included in this document reflects only the authors' views. The European Community is not liable for any use that may be made of the information contained herein.